# Full-text and URL search

Miguel Costa

**Abstract** Web archives are a historically valuable source of information. In some respects, web archives are the only record of the evolution of human society in the last two decades. They preserve a mix of personal and collective memories, the importance of which tends to grow as they age. However, the value of web archives depends on their users being able to search and access the information they require in efficient and effective ways. Without the possibility of exploring and exploiting the archived contents, web archives are useless. Web archive access functionalities range from basic browsing to advanced search and analytical services, accessed through user-friendly interfaces. Full-text and URL search have become the predominant and preferred forms of information discovery in web archives, fulfilling user needs and supporting search APIs that feed complex applications. Both full-text and URL search are based on the technology developed for modern web search engines, since the Web is the main resource targeted by both systems. However, while web search engines enable searching over the most recent web snapshot, web archives enable searching over multiple snapshots from the past. This means that web archives have to deal with a temporal dimension that is the cause of new challenges and opportunities, discussed throughout this chapter.

Miguel Costa
Vodafone Research, e-mail: miguel.costa2@vodafone.com

# 1 Introduction

The World Wide Web has a democratic character, and everyone can publish all kinds of information using different types of media. News, blogs, wikis, encyclopedias, photos, interviews and public opinion pieces are just a few examples. Some of this information is unique and historically valuable. For instance, the online newspapers reporting the speech of a president after winning an election or announcing an imminent invasion of a foreign country might become as valuable in the future as ancient manuscripts are today. Historical interest in these documents is also growing as they age, becoming a unique source of past information for many areas, such as sociology, history, anthropology, politics, journalism, linguistics or marketing.

Much of the current effort concerned with web archive research and development focuses on acquiring, storing, managing and preserving data (Masanès, 2006). However, to make historical analysis possible, web archives must turn from mere document repositories into accessible archives. Full-text search, i.e., finding documents with text that match the given keywords or sentences, has become the dominant form of information access, especially thanks to web search engines such as Google, which have a strong influence on how users search in other systems. In web archives, full-text queries (e.g., Iraq war) can be narrowed to a user-specified time interval (e.g., year 2003). The matching document versions, sometimes in the order of millions, are then ranked according to their relevance to the query and time interval. Surveys indicate that full-text search is the preferred method for accessing web archive data, and the most used when supported (Costa and Silva, 2011; Ras and van Bussel, 2007). As a result, even with the high computational resources required to provide full-text search over large-scale web collections, 63% of worldwide web archives provide it for at least a part of their collections (Costa et al., 2016).

As a result of the challenges of creating effective and efficient mechanisms to match and rank all document versions, the prevalent discovery method in web archives is based on URL (Uniform Resource Locator) search. Users can search for a web document of interest by submitting its URL and narrow the document versions by date range. A list of chronologically ordered versions is then returned for that URL, such as in the Internet Archive's Wayback Machine[1]. This allows, for instance, a search for stored versions of a website by using its original web address. Around 80% of worldwide web archives support this type of search, which requires much less computational resource than full-text search. On the other hand, URL search is limited, as it forces users to remember URLs which may have ceased to exist a long time ago.

Metadata search, i.e., a search by metadata attributes, such as category, language and file format, is another type of search that complements the previous two examples. For instance, the Library of Congress Web Archive[2] supports search on the bibliographic records of its collections. Other web archives include metadata filters in their advanced search user interfaces along with full-text or URL search. The manual creation of metadata describing the curated collections and their artefacts is a time-consuming and expensive process, which makes it a non-viable option for large-scale web archives. In this case, most of the metadata must be created automatically. Metadata search is provided by 72% of worldwide web archives.

---

[1] http://web.archive.org

[2] http://www.loc.gov/webarchiving

These three types of search are available in most web archives, complementing each other. Together, full-text, URL and metadata search are the only form of information discovery in web archives, with the exception of some prototypes. The technology is adapted from the typical search functionalities provided by modern web search engines. The technology developed to enable search over the most recent web snapshot was extended to enable search over multiple snapshots from the past. The main advantage of this solution is that people are used to these search functionalities.The main disadvantage is that this technology tends to provide unsatisfactory results, because the data characteristics of web archives and the information needs of their users are different (Costa and Silva, 2010). Hence, the web archiving community have been working on search tools and user interfaces that better satisfy web archive users.

The remainder of this chapter is organised as follows. Section 2 presents the typical search user interfaces offered by web archives, giving an overview of web archive access functionalities. Section 3 characterises why, what and how users search, which supports technology development decisions and suggests new research directions. Section 4 explains how current search technology locates the desired information and fulfils users' information needs, and Section 5 draws together the conclusions of this research.

## 2 Search user interfaces

Most web archives are accessible from the live Web through a web browser. Their digital collections, to which access may be restricted, are searchable via a graphical user interface (GUI). Despite the particularities and different layouts, these GUIs generally provide similar functionalities. The GUIs are mostly composed of a search box complemented with a date range filter to narrow the search results to a specific time interval. The GUIs of the three web archives depicted in this Section are illustrative of this arrangement, namely, the Internet Archive[3], the Portuguese Web Archive[4] and the Library of Congress Web Archive[5]. The Internet Archive, created in 1996, was one of the first web archives and leads the most ambitious initiative, with more than 400 billion web pages saved worldwide over time. Its Wayback Machine is probably the most used system for searching and accessing archived content, along with its open-source counterpart OpenWayback[6], used by most web archives. Because of the huge amount of data, full-text search is only available in beta mode for a few million documents[7]. Currently, URL search is the main method to find information in the Internet Archive.

---

[3] http://archive.org

[4] http://archive.pt

[5] http://loc.gov/websites

[6] http://github.com/iipc/openwayback

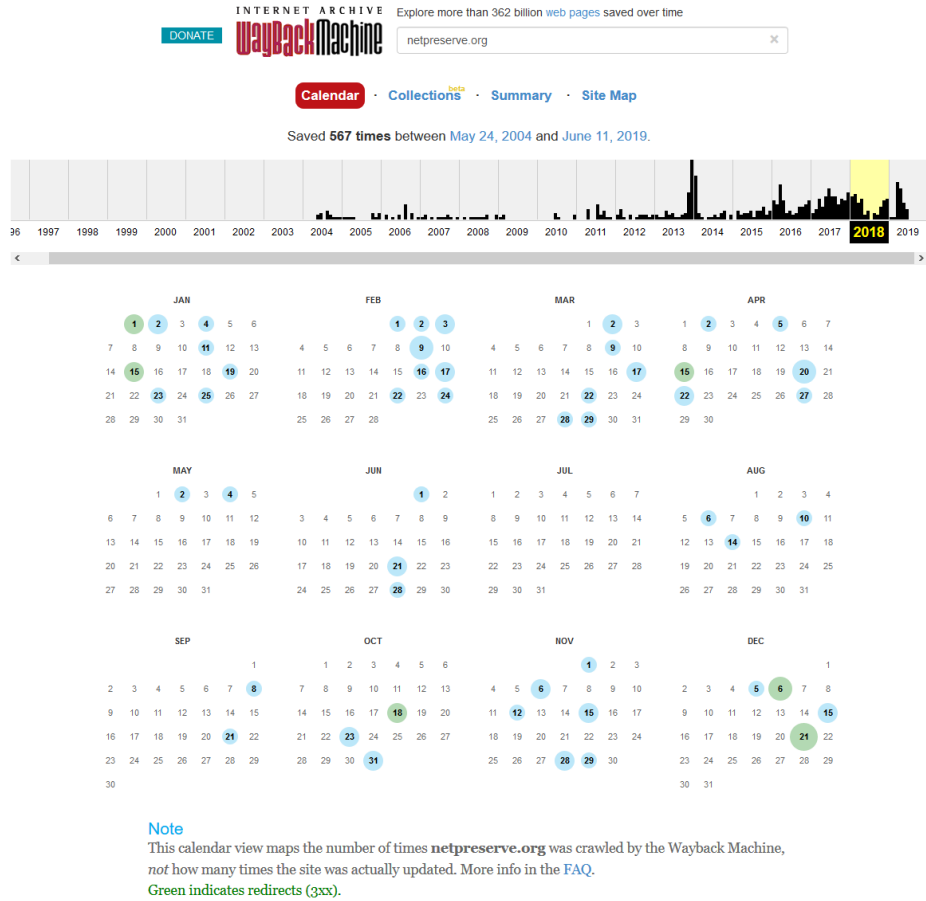

**Fig. 1** User interface of the Internet Archive's Wayback Machine after a URL search (*netpreserve.org* was submitted and 2018 selected to narrow the archived versions to this year).

Fig. 1 depicts the GUI of the Wayback Machine after searching a URL without protocol (e.g. http), which in this example is *netpreserve.org*. A calendar view is presented with circles around the days when the relevant web page was archived. The size of the circle indicates how many versions exist on a specific day. A total of 567 versions were archived between 2004 and 2019. A time slider at the top of the GUI shows the distribution of the number of archived versions throughout the years and enables the user to select a year and narrows the calendar view to that period.

---

[7] http://blog.archive.org/2016/10/26/searching-through-everything/

**Fig. 2** User interface of the Portuguese Web Archive after a full-text search (only the top two results are shown for the query *jornal record* which means *record newspaper* in Portuguese, and the archived versions are restricted to between 1996 and 2018).

In this example, the selected year is 2018. The user can then click on the versions to view and navigate them as they were in the past.

The Memento project adds a temporal dimension to the HTTP protocol so that archived versions of a document can be served by the web server holding that document or by existing web archives that support the Memento API (application programming interface), such as the Internet Archive (Van de Sompel et al., 2009). The Time Travel portal[8] uses this protocol to propagate URL queries to the aforementioned systems and redirects users to the archives which host the versions of a given URL. For each archive, the version closest to the requested date is shown, along with the first, last, previous and next versions. The portal works as a metasearch engine for web archives.

The Portuguese Web Archive (PWA) was created in 2008 and preserves the Portuguese web, which is considered the web of most interest for the Portuguese community. It provides access to more than 6.5 billion documents, searchable by full-text and URL via the same text box. The GUI of the PWA automatically interprets the type of query and presents the results accordingly. When the PWA receives a full-text query, which can be restricted by a date range, it returns a results page containing a list of 10 web pages matching the query and the time interval, as illustrated in the example of Fig. 2. User interaction with the system and the results layout are similar to commercial web search engines. The results are ranked by relevance to the query, determined by the PWA ranking model. Each result includes the title, URL and crawled date of the web page, along with a snippet of text containing the query terms. The user can then click on the links of the archived web pages to view and navigate them. In addition, the user can click to see a list of all archived versions chronologically ordered for each web page.

---

[8] http://timetravel.mementoweb.org

**Fig. 3** User interface of the Library of Congress Web Archive after a keyword search on metadata (*elections* was submitted to search on the indexed *Archived Web Sites*).

Both the Wayback Machine and the PWA offer an advanced interface for users to conduct metadata search by applying multiple filters (e.g. file format, domain name) or by restricting the search to special operators (e.g. exact phrases, word exclusion). Another example of metadata search is the Library of Congress Web Archive, which was set up in 2000 and since then has been creating collections of archived websites on selected topics, such as the U.S. elections or the Iraq War. It enables users to search for keywords in the bibliographic records of their collections and then to refine results by multiple filters, such as format type, location, subject, language and access conditions. Fig. 3 shows the GUI of the Library of Congress Web Archive with some of these filters available on the left-hand side.

In addition to GUIs, several APIs implemented as web services have been provided to facilitate interoperability between web archives and external applications. These APIs specify how to search and access web archive collections automatically.

Examples include the Internet Archive API[9], the Arquivo.pt API[10] and the Memento Time Travel API[11]. They are an essential piece of software to feed complex applications and GUIs with historical data.

## 3 Why, what and how do users search?

Search is a way to achieve users' goals and not an end in itself. Understanding these goals along with all other aspects of user behaviour is key to the success of web archives. This knowledge enables the development of technology, the design of GUIs and the tailoring of search results that better satisfy users. To this end, a few studies have been conducted to understand why, what and how users search in web archives.

Let me first introduce the taxonomy proposed by Broder, which is common to information retrieval researchers and used to investigate different search systems, of which web search engines are the most studied (Broder, 2002). Broder classified web search engine queries into three broad categories according to user goal: 1) *navigational* to reach a particular web page or site (e.g., archive.org); 2) *informational* to collect information about a topic, usually from multiple pages, without having a specific one in mind (e.g., Brexit); 3) *transactional* to perform a web-mediated activity (e.g., shopping, downloading a file, finding a map).

Quantitative and qualitative studies investigated the information needs of general web archive users (i.e., why they are searching). Three studies presented in the same research article, namely, a search log analysis, an online questionnaire and a laboratory study, found similar results (Costa and Silva, 2010). The information needs of web archive users are mainly navigational, then informational and lastly, transactional. Overall, these needs are aligned with the available functionalities. URL search is a perfect match to fulfil navigational needs when URLs are submitted, which occurs on a significant percentage of occasions (26% reported in (Costa and Silva, 2011)). Additionally, full-text search can be used as a generic tool to fulfil all three needs. In sum, the search technology provided by web archives can support the main informational needs of their users. On the other hand, web archives fail in supporting some specific needs, such as exploring the evolution of a website.

Full-text and URL search are also limited when analysing complex information needs, such as those of historians or journalists. For instance, they fail to provide answers to questions such as "Who were the most discussed politicians during 2005?", "What were the top sporting events in the last decade?" or "What were the most popular time periods related to Barack Obama?". Users would have to submit many queries without guarantees of full coverage of all relevant content. The search functionalities of web archives lack expressive power to query an entity (e.g., a person) or an event (e.g., September 11) by its properties or relations with other entities and events (Fafalios et al., 2018). Researchers find it difficult to meet these complex information needs when they use web archives as a source of research data (Brügger and Milligan, 2018; Dougherty et al., 2010; Singh et al., 2016b).

---

[9] http://archive.org/services/docs/api
[10] http://github.com/arquivo/pwa-technologies/wiki/APIs
[11] http://timetravel.mementoweb.org/guide/api

When we look at what users search for in order to understand what is relevant for them, results show that nearly half of their informational needs are focused on names of people, places or things. Many navigational queries only contain the names of companies or institutions. People tend to remember short names that they use as keywords for searching. When analysing by topic, navigational queries refer mostly to websites about commerce, computers or the Internet, such as blogs, and about education, such as universities. The informational queries refer mostly to people, health and entertainment. This search analysis was conducted on the general use of the Portuguese Web Archive (Costa and Silva, 2011). Another study on the Wayback Machine indicates that its users mostly request English pages, followed by pages written in European languages (AlNoamany et al., 2014). Most of these users probably search for pages archived in the Wayback Machine because the requested pages no longer exist on the live Web.

Studies about the ways in which users search indicate that they do not spend much time and effort searching the past Web (AlNoamany et al., 2013; Costa and Silva, 2011). Users prefer short sessions, composed of short queries and few clicks. Full-text search is preferred to URL search, but both are frequently used. This preference is understandable because URL search requires users to remember the exact URL, which in some cases became unavailable a long time ago. User interactions in the form of queries, query terms, clicked rankings and viewed archived pages follow a power law distribution. This means that all these interactions have a small percentage that is repeated many times and can be exploited to increase the performance of web archives (e.g., using cache mechanisms) and the quality of their results (e.g., learning a higher ranking for the most viewed pages). There is a strong evidence that users prefer the oldest documents over the newest, but mostly do not search with any temporal restriction or use temporal expressions. This is surprising, since all information needs are focused on the past and not even large time intervals were used to narrow queries. Also surprising is the fact that web archive users said they wanted to see the evolution of a page over time, but they tended to click on just one or two versions of each URL. Web archive users search as they would in web search engines. These behaviours may be the result of offering similar GUIs, leading users to search in a similar way. Another example of this behaviour is the fact is that almost all users of both web archives and web search engines do not use metadata filters.

Lastly, specific types of researchers, such as historians, present different search behaviours (Singh et al., 2016b). Since they require a comprehensive understanding of a subject, they usually start with broader queries about that subject, which are reformulated several times to get an overview of the areas of interest. More specific queries are then submitted on each research aspect in order to gain deeper knowledge.

## 4 Finding the right information

Finding the desired information is a complex process that passes through several stages. This section explain them very succinctly and gives a glimpse of the associated challenges.

### 4.1 From archived data to search results

URL search matches all archived versions of a given URL and its possible aliases (i.e., different URLs referring to the same resource). URLs are first normalised to a canonical form and then indexed to speed up the matching phase. For instance, *http://www.example.com*, *http://example.com:80* and *http://example.com/index.html* tend to refer to the same web page and can be normalised to *http://example.com*. The matching versions are traditionally presented in a chronological view, such as shown in Fig. 1. The versions are ranked by their timestamp, which usually refers to the content acquisition date, but may also refer to the creation or publication date. This ranking enables users easily to jump between the oldest and most recent versions.

URL search can be implemented without processing the full content of archived documents, using for instance just the crawl logs. Creating a full-text search service is much more complex. All data must be processed beforehand, for example broken up into words (tokenised) and syntactically analysed (parsed) to separate text from metadata and identify the structural elements to which each segment of text belongs (e.g. title, headings, image). This processing is performed in hundreds of file formats that continue to evolve, such as HTML, PDF and Microsoft Office formats. Further processing is usually carried out, such as link extraction for link analysis algorithms (e.g. PageRank (Page et al., 1998)) and the enhancement, with anchor text, of the content of web documents to which the links point. Then, index structures are created over the words and metadata to speed up the matching process between documents and queries. This matching depends on the implemented retrieval model. Usually, for very large-scale collections such as the Web, a retrieval model is chosen, where all query terms or semantically related terms must occur in the documents. This works as a selection stage for candidate documents relevant to the query. Still, millions of documents may be retrieved, which makes it extremely hard for a user to explore and find relevant information. To overcome this problem, ranking models estimate document relevance based on how well documents match user queries. Documents are then sorted in descending order by their relevance score as a means for users to find information effectively and efficiently.

### 4.2 Ranking search results

A ranking model combines a set of ranking features (a.k.a, factors or signals). Some of these features estimate the documents' relevance according to a given query. Examples include query term proximity in documents, where closer terms tend to be more related than terms which are further apart, or the TF-IDF and BM25 functions (Robertson et al., 2009), which measure the relevance of query terms to a document based on the number of times the terms occur in the document and collection. Others features are independent from the query and estimate a measure of importance, quality or popularity for a document. Examples include the number of links a document receives or the number of times a document was visited. There are many other proven ranking features that can be used alone or in a combination (Baeza-Yates and Ribeiro-Neto, 2011).

Creating precise ranking models for a new type of search, such as for web archives, brings many challenges. First, it requires a comprehensive knowledge of users, as described in Section 3. Second, it is necessary to quantify user relevance with the proper ranking features, some of which need to be created or redesigned for the new context. Third, all features need to be combined into one ranking model optimised towards a goal (e.g., an evaluation metric aligned with user relevance). Combining them manually is not trivial, especially when there are hundreds of potential features. Additionally, manual tuning can lead to outfitting, i.e., the model fits training data closely, but fails to generalise to unseen test data. Hence, supervised learning-to-rank (L2R) algorithms have been employed to automatically find the best way to combine ranking features, resulting in significant improvements (Liu, 2009). Deep learning extends L2R capabilities with novel algorithms that can automatically learn representations of queries and documents for matching in a latent semantic space (Mitra and Craswell, 2018). Additionally, end-to-end ranking models can be learned directly from data without the need to build hand-crafted features or even manually labelling datasets.

### 4.3 Web archive information retrieval

All steps necessary to provide ranked search results require significant time and effort. This is the reason why most web archives use web search engine technology to support full-text search on their collections. Most of this technology uses Lucene[12] or its extensions for distributed processing, such as Solr[13], ElasticSearch[14] and Nutch-WAX[15]. All these ignore the specificities of web archives, and the consequence is that the search results generated are poor quality and fail to satisfy users (Costa et al., 2014).

---

[12] http://lucene.apache.org

[13] http://lucene.apache.org/solr

[14] http://www.elastic.co

[15] http://archive-access.sourceforge.net

Web archive information retrieval is a recent field which is still developing solutions for web archive users. As a result, some of the approaches have come from the field of temporal information retrieval, which also considers both topical and temporal criteria of relevance (Campos et al., 2015; Kanhabua et al., 2015). An idea shared between both fields is that the distribution of document dates can be exploited, since it reveals time intervals that are likely to be of interest to the query (Jones and Diaz, 2007; Singh et al., 2016b). For instance, when searching for *tsunami*, the peaks in the distribution may indicate when tsunamis occurred. Another idea is to favour more dynamic documents, since documents with higher relevance are more likely to change or change to a greater degree (Elsas and Dumais, 2010). More popular and revisited documents are also more likely to change (Adar et al., 2009). On the other hand, the most persistent terms are descriptive of the main topic and are probably added early in the life of a document. These persistent terms are especially useful for matching navigational queries, because the relevance of documents for these queries is not expected to change over time.

This type of knowledge about what is relevant for users is core for the engineering of better ranking features for web archives. A different type of improvement can be achieved by training temporal ranking models. For instance, multiple models trained and optimised for specific time periods showed better performance than using just one single model to fit all data (Costa et al., 2014). A simpler way to take advantage of multiple models is to allow users to select the one that is better suited for their particular information need (Singh et al., 2016a). Scholars, for instance, initially want an overview of the research topic, preferring more diversified results, and later on in the process they need more focused results. Different models can support these different needs.

### 4.4 Scalability challenges

Creating a full-text search service on the huge and rapidly growing amount of archived data presents many scalability challenges. Users expect Google response times and optimal search results, but the web archiving community has only a very small percentage of the budget of commercial web search engines. Thus, some creative alternatives have been proposed, such as simply indexing the anchor texts, which are many orders of magnitude smaller (Holzmann et al., 2017). Anchor texts to a web document are short and concise descriptions that together represent the collective perception of other content authors.

Besides anchors, URLs may contain hints about the content of documents, such as their language (Baykan et al., 2008), topic (e.g., sport) (Baykan et al., 2009) and genre (e.g., blog) (Abramson and Aha, 2012). Named entities can be extracted from URLs to annotate documents accurately (Souza et al., 2015). Combined, the non-content features provide better search results than individually, which may include metadata extracted from crawl logs (e.g., number of archived versions), hyperlinks (e.g., number of inlinks), URL strings (e.g., URL depth) and anchors (e.g., frequency of the query terms in the anchor text) (Vo et al., 2016).

A completely different approach to tackling the scalability challenges is to use commercial web search engines (e.g., Bing) to retrieve lists of ranked results from the live Web (Kanhabua et al., 2016). Results are then linked to the Wayback Machine to support browsing in the archived versions. This solution supports full-text search almost without developing any technology. The downside is that only a part of the archived web is covered by commercial web search engines, and this tends to worsen over time as more pages become permanently unavailable. Moreover, this technology is not tailored for web archive users. For instance, it is known that web search engines favour fresh over old content, which does not apply to web archives.

## 5 Conclusions

Full-text and URL search turned web archives into useful data sources for a wide range of users and applications. Currently, these forms of search are the best and most available forms of information discovery in web archives. They are an amazing feat of engineering, able to process billions of archived document versions to enable discovery of the desired information almost immediately. Nevertheless, web archives continue to adopt web search engine technology that ignores the particularities of their users and preserved collections. As a result, the quality of search results, and thus user satisfaction, is far from ideal.

The web archiving community has offered some improvements, such as better GUIs, search tools and ranking algorithms. One of the reasons for these improvement is a better understanding of the users, such as knowing why are they searching, what is relevant for them and how they search to fulfil their information needs. Another reason is the joint efforts of the community to develop common tools and data formats for web archives. These are promising advances to make digital memories easier to find and exploit.